\documentclass[showpacs,superscriptaddress,floatfix,aps,prl,notitlepage,reprint]{revtex4-2}
\usepackage{graphicx,amsfonts,amssymb,amsmath,enumerate}
\usepackage[hidelinks]{hyperref}
\usepackage{hypcap}
\usepackage{braket}
\usepackage{amsthm}
\usepackage{multirow}
\usepackage{dcolumn}   %
\usepackage{bm}        %
\usepackage{epstopdf}
\usepackage{mathtools}

\usepackage{color}
\usepackage{subfigure}
\usepackage{diagbox}
\usepackage{longtable}
\usepackage{makecell}
\usepackage{bbold}
\usepackage{algorithm}
\usepackage{algpseudocode}
\usepackage{comment}
\usepackage{svg}

\DeclareMathAlphabet{\mathitb}{OT1}{cmr}{bx}{sl}
\usepackage[utf8]{inputenc}
\usepackage{enumitem}

\newcommand{\mN}{\mathcal{N}}
\newcommand{\mO}{\mathcal{O}}
\newcommand{\mF}{\mathbb{F}}
\newcommand{\mE}{\mathbb{E}}
\newcommand{\mC}{\mathcal{C}}

\usepackage{etoolbox}
\makeatletter
\patchcmd{\frontmatter@abstract@produce}
  {\vskip200\p@\@plus1fil
   \penalty-200\relax
   \vskip-200\p@\@plus-1fil}
  {}
  {}
  {}
\makeatother

\begin{document}

\title{Exponentially Accelerated Sampling of Pauli Strings for Nonstabilizerness}

\author{Zhenyu Xiao}
\email{zyxiao@princeton.edu}
\affiliation{Princeton Quantum Initiative, Princeton University, Princeton, New Jersey 08544, USA}

\author{Shinsei Ryu}
\affiliation{Department of Physics, Princeton University, Princeton, New Jersey 08544, USA}

\date{\today}

\begin{abstract}
Quantum magic, quantified by nonstabilizerness, measures departures from stabilizer structure and underlies potential quantum speedups.
We introduce an efficient classical framework for computing stabilizer R\'enyi entropies and stabilizer nullity of generic $N$-qubit wavefunctions.
The method combines the fast Walsh-Hadamard transform with an exact partition of Pauli operators, reducing the average cost per sampled Pauli string from $\mathcal{O}(2^N)$ to $\mathcal{O}(N)$.
We further develop a Monte Carlo estimator with Clifford preconditioning and find that the required number of samples shows no visible growth with $N$ in our benchmarks. 
Applying the method to $T$-doped random Clifford circuits, we identify the scrambling ratio $\eta$ (Clifford gates per $T$ gate) as the key parameter governing magic growth.
Each $T$ gate approaches its dilute-limit nonstabilizerness power with only modest Clifford scrambling.
Our approach enables quantitative studies of magic
in highly entangled states and long-time nonequilibrium dynamics.
\end{abstract}

\maketitle

{\it Introduction.---}
Quantum computers are exponentially faster than classical ones for certain computational tasks~\cite{shor1997,childs2010}. 
Such quantum advantage relies on distinctive features of quantum states, including entanglement~\cite{amico2008,eisert2010,cirac2012,nielsen2010}, yet entanglement alone is not sufficient.
Clifford circuits acting on stabilizer states can generate extensive entanglement but still be efficiently simulated classically~\cite{aaronson2004,gottesman1998}.
Therefore, genuine quantum speedup requires non-Clifford operations to generate non-stabilizer states.
This motivates the notion of nonstabilizerness, also known as quantum magic~\cite{bravyi2005,howard2014,veitch2014}. 
It quantifies departures from stabilizer structure and, within a resource-theoretic viewpoint, characterizes the difficulty of preparing the state~\cite{veitch2014,chitambar2019}.
Understanding how nonstabilizerness is generated and redistributed across different quantum dynamics~\cite{wang2020a,fux2024a,bejan2024,bejan2025,zhang2024,turkeshi2025,smith2025,odavic2023,lami2025,falcao2025b,aditya2025,aditya2025a,dowling2025a,hou2025,szombathy2025a,maity2025,sticlet2025,passarelli2025,magni2025,aziz2025,tarabunga2025,scocco2025,campbell2010,trigueros2025,rattacaso2023,niroula2024,paviglianiti2025,liu2022} is important in quantum science and engineering.

Beyond its role as a computational resource, nonstabilizerness has emerged as a diagnostic for many-body physics, including phase transitions~\cite{oliviero2022,tarabunga2024b,falcao2025a}, conformal field theory~\cite{white2021a,qian2025a,frau2025,hoshino2025a}, quantum chaos~\cite{leone2021a,turkeshi2025a,jasser2025,zhang2025f,bera2025a}, and thermalization~\cite{tirrito2025a,odavic2025,turkeshi2025,falcao2025b,nandkishore2015,liu2022,korbany2025a,santra2025,haug2025}. 
However, quantifying nonstabilizerness typically involves nonlinear functionals of the many-body wave function and is therefore notoriously difficult in practice~\cite{liu2022}.

There exist several measures of nonstabilizerness in quantum information theory, such as the robustness of magic and the relative entropy of magic~\cite{howard2014,beverland2020,liu2022}. 
These quantities are defined through optimizations over operator decompositions, making direct numerical evaluation impractical beyond a few qubits~\cite{heinrich2019}. 
More recently, computable diagnostics based on the Pauli expansion, $P\in \{I,X,Y,Z \}^{\otimes N}$, have been introduced, including stabilizer R\'enyi entropy~\cite{leone2022}, stabilizer nullity~\cite{beverland2020,jiang2023d}, and Bell magic~\cite{haug2023c}.
While these Pauli-string-based measures avoid explicit optimizations, their numerical cost remains substantial. 
For a generic $N$-qubit state $|\psi\rangle$ represented as a full state vector, evaluating a single correlator $\langle\psi|P|\psi\rangle$ requires $\mO(2^N)$ time. 
Consequently, a brute-force evaluation that enumerates all $2^{2N}$ Pauli strings scales as $\mO(2^{3N})$.

This difficulty is alleviated when the state admits an efficient classical representation, e.g., as a matrix product state~\cite{lami2024,chen2024a,haug2023a,tarabunga2024a,frau2024b}.
However, for long-time dynamics, generic states develop volume-law entanglement, and the required bond dimension and cost grow exponentially with $N$.
Monte Carlo (MC) sampling over Pauli strings provides an alternative.
For full state vectors, sampling $\mN$ Pauli strings by direct evaluation costs $\mO(\mN 2^N)$.
Perfect-sampling schemes have been introduced with cost $\mO(\mN 2^{\frac{3}{2}N})$ for full states~\cite{tarabunga2025a} and $\mO(\mN N\chi^3)$ for matrix product states with bond dimension $\chi$~\cite{lami2023}.
In practice, sampling can still be hard when the Pauli-weight distribution is strongly inhomogeneous, where $\mN$ may need to grow exponentially with $N$.
Finally, specialized methods exist for restricted families of states~\cite{collura2025b, wang2025,ding2025,liu2025c,hinsche2025}, but efficient approaches for generic many-body states remain limited.

In this Letter, we develop an efficient classical framework for computing Pauli-string-based measures of nonstabilizerness for generic $N$-qubit wavefunctions.
We partition Pauli operators into $2^N$ families $\{P_{x,z}\}_{z\in\mF_2^N}$ and evaluate all correlators within each family simultaneously via a fast Walsh-Hadamard transform (FWHT)~\cite{shanks1969}.
FWHT has been used to estimate Pauli channels~\cite{chen2023,flammia2020,harper2021} and decompose matrices~\cite{georges2025}, but its potential for evaluating correlators of quantum many-body states remains unexplored.
Within this framework, the cost of obtaining stabilizer R\'enyi entropy and stabilizer nullity is reduced from brute-force $\mO(2^{3N})$ to $\mO(N2^{2N})$.
We further introduce a MC estimator with Clifford preconditioning, after which the required sample number shows no visible growth with $N$ in our benchmarks.
We then apply the method to study magic generation in $T$-doped random Clifford circuits. 
Clifford scrambling between non-Clifford gates~\cite{fux2024a,gu2025,zhou2020c,bejan2024,bejan2025} is known to be crucial for fully realizing their nonstabilizerness power~\cite{szombathy2025b,varikuti2026,hou2025}, 
but how much scrambling suffices has remained unclear.
We identify the scrambling ratio $\eta$ (Clifford gates per $T$ gate) as the governing parameter and find that its effect on the growth rate of stabilizer R\'enyi entropy saturates at $\eta \gtrsim 5$.

{\it Stabilizer R\'enyi entropy and stabilizer nullity.}---
We consider an $N$-qubit pure state $|\psi\rangle$ with $d=2^N$.
Expanding the density matrix in the Pauli basis $P\in\{I,X,Y,Z\}^{\otimes N}$, we write $|\psi\rangle\!\langle\psi|=\sum_{P}\frac{1}{ d}\,c_P\,P$ with
$c_P=\langle\psi|P|\psi\rangle$.
Using Pauli orthogonality and $\mathrm{Tr}[(|\psi\rangle\!\langle\psi|)^2]=1$, one has
$\sum_P \frac{1}{d}|c_P|^2=1$, so $\{\frac{1}{d}|c_P|^2\}$ defines a probability distribution over Pauli strings.
The stabilizer R\'enyi entropy is the R\'enyi entropy of this distribution (shifted by $N$)~\cite{leone2022},
\begin{equation}
M_\alpha(|\psi\rangle)
:=\frac{1}{1-\alpha}\log_2\!\left(\sum_P {|c_P|^{2\alpha}}/{d^{\alpha}}\right)-N .
\label{eq:SRE-def}
\end{equation}
The stabilizer nullity~\cite{beverland2020,jiang2023d} is defined by
\begin{equation}
\nu(|\psi\rangle)
:=N-\log_2\big|{\rm STAB}(|\psi\rangle)\big|,
\end{equation}
where ${\rm STAB}(|\psi\rangle)$ denotes the stabilizer group of $|\psi\rangle$ and
$\big|{\rm STAB}(|\psi\rangle)\big|$ counts Pauli strings $P$ satisfying
$P|\psi\rangle=\pm|\psi\rangle$.
Both $M_\alpha(|\psi\rangle)$ and $\nu(|\psi\rangle)$ are non-negative, vanish if and only if $|\psi\rangle$ is a stabilizer state, and are invariant under Clifford operations~\cite{leone2022}.
In magic-state resource theory, $M_\alpha$ is not a monotone for $\alpha<2$~\cite{haug2023b}, whereas $M_\alpha$ for $\alpha\geq2$ and $\nu$ are monotones for pure states under stabilizer operations~\cite{beverland2020,leone2024}.

{\it Fast Walsh-Hadamard Pauli sampling.---}
To obtain numerically exact values of $M_\alpha(|\psi\rangle)$ or $\nu(|\psi\rangle)$ by brute-force Pauli enumeration, one would in general need all $d^2$ correlators $\langle\psi|P|\psi\rangle$.
We first illustrate the key idea by restricting it to Pauli strings containing only $I$ and $Z$:
$P_z:=Z^{z_1}\otimes Z^{z_2}\otimes\cdots\otimes Z^{z_N}$ with $z_i\in\{0,1\}$ and $z=(z_1,\dots,z_N)\in\mF_2^N$.
Writing $|\psi\rangle=\sum_{b\in\mF_2^N}\psi(b)|b\rangle$, we have $P_z|b\rangle=(-1)^{z\cdot b}|b\rangle$ with $z\cdot b:=\sum_i z_i b_i$, 
and hence
\begin{equation}
\langle\psi|P_z|\psi\rangle=\sum_{b\in\mF_2^N}(-1)^{z\cdot b}\,|\psi(b)|^2,
\label{eq:Z-FWHT}
\end{equation}
which is precisely the discrete Fourier transform on $\mF_2^N$ 
(Walsh-Hadamard transform) of $f(b):=|\psi(b)|^2$~\cite{walsh1923}.
Computing all $\{\langle\psi|P_z|\psi\rangle\}_z$ by brute force costs $\mO(2^{2N})$, whereas the fast Walsh-Hadamard transform (FWHT) evaluates the full transform in $\mO(N2^N)$ time via a hierarchy of pairwise sum-difference updates~\cite{shanks1969}.
The FWHT can be viewed as iteratively applying the Hadamard gate
$H=\frac{1}{\sqrt2}\begin{pmatrix}1&1\\1&-1\end{pmatrix}$
to the length-$2^N$ vector $|f\rangle:=\sum_{b\in\mF_2^N} f(b)\,|b\rangle$.
Applying $\sqrt2 H$ to the first qubit yields
\begin{equation}
\sqrt2 H|f\rangle
=\sum_{z_1,b_{\bar1}}\left(\sum_{b_1\in\mF_2}(-1)^{z_1b_1}\,f(b_1,b_{\bar1})\right)\,|z_1,b_{\bar1}\rangle,
\end{equation}
with $b=(b_1,b_{\bar1})$.
This step is the $\mF_2$ Fourier transform on the first index of $f(b_1,b_{\bar1})$.
Iterating over all qubits yields
$(\sqrt2 H)^{\otimes N}|f\rangle=\sum_{z\in\mF_2^N}\langle\psi|P_z|\psi\rangle\,|z\rangle$,
so $\{\langle\psi|P_z|\psi\rangle\}_z$ are read off from the amplitudes.
Each application of $\sqrt2 H$ performs $\mO(2^N)$ floating-point operations, giving an overall time complexity of $\mO(N2^N)$.

To generalize this idea to arbitrary Pauli strings, we use a binary labeling.
Any $N$-qubit Pauli can be written as a product of $X$ and $Z$ operators with an overall phase, and is specified by a pair
$(x,z)\in\mF_2^N\times\mF_2^N$~\cite{gottesman1998,nielsen2010},
\begin{equation}
P_{x,z}=e^{i\phi(x,z)}\,X^x Z^z,\,
X^x:=\bigotimes_{j=1}^N X_j^{x_j},\,
Z^z:=\bigotimes_{j=1}^N Z_j^{z_j},
\label{eq:Pxz-def}
\end{equation}
where the phase $e^{i\phi(x,z)}$ does not affect $|\langle\psi|P_{x,z}|\psi\rangle|$.
In the computational basis $\{|b\rangle\}_{b\in\mF_2^N}$, $X^x$ acts as a bit flip, $X^x|b\rangle=|b\oplus x\rangle$, while $Z^z|b\rangle=(-1)^{z\cdot b}|b\rangle$.
Therefore, for $|\psi\rangle=\sum_b \psi(b)\,|b\rangle$,
\begin{equation}
\langle\psi|P_{x,z}|\psi\rangle
=e^{i\phi(x,z)}\sum_{b\in\mF_2^N}\overline{\psi(b)}\,(-1)^{z\cdot b}\,\psi(b\oplus x).
\label{eq:Pxz-corr}
\end{equation}
For each fixed $x$, define $f_x(b):=\overline{\psi(b)}\,\psi(b\oplus x)$.
Equation~\eqref{eq:Pxz-corr} is exactly the Walsh-Hadamard transform of $f_x$ from $b$-space to $z$-space, so a single FWHT produces $\{\langle\psi|P_{x,z}|\psi\rangle\}_{z\in\mF_2^N}$ in $\mO(N2^N)$ time.
Sweeping $x$ over $\mF_2^N$ enumerates the expectation values of all $4^N$ Pauli strings, for a total cost of $2^N$ FWHTs, i.e.\ $\mO(N2^{2N})$ time complexity.
This yields an exponential speedup compared to brute-force enumeration, which scales as $\mO(2^{3N})$ when each correlator is evaluated in $\mO(2^N)$ time.

\begin{figure}[tbp]
  \centering
  \includegraphics[width=0.47\textwidth]{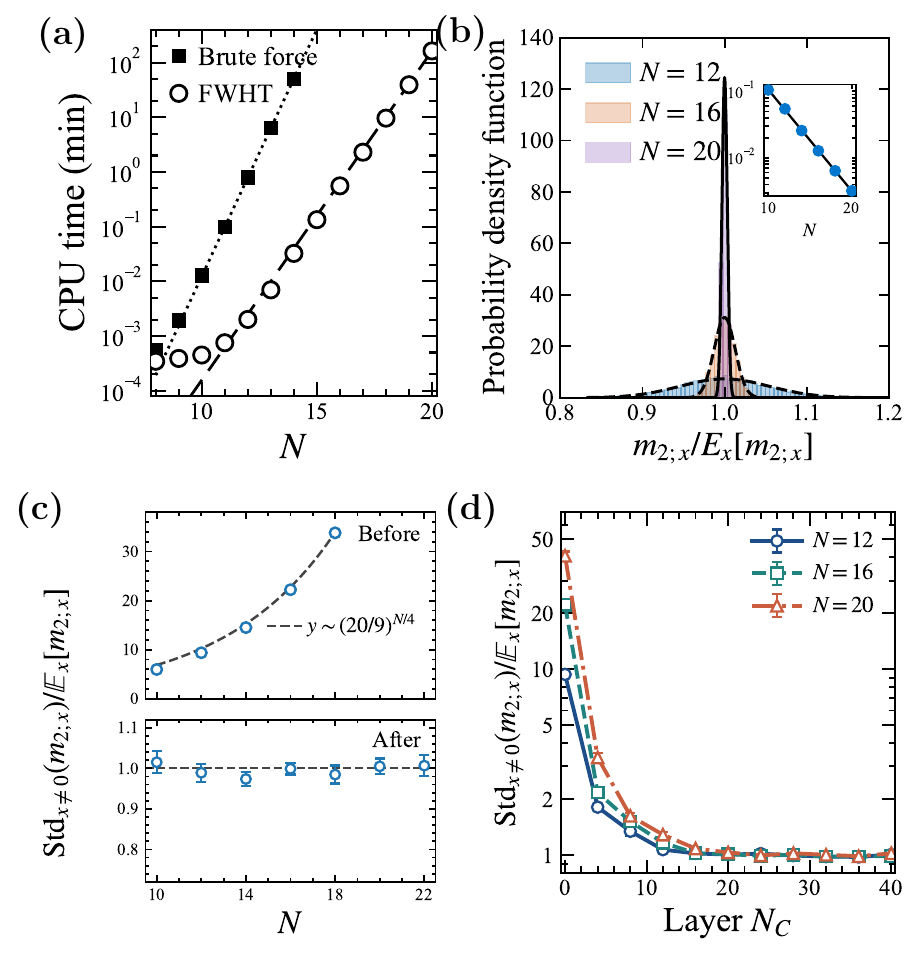}
\caption{(a) Computation time for $M_2$ of random magic states by brute-force versus FWHT-based Pauli enumeration. Dashed lines indicate $\mO(2^{3N})$ and $\mO(2^{2N})$ scaling.
(b) Probability density of $m_{2;x}/\mE_x[m_{2;x}]$ for Haar-random states (black: Gaussian fits; $x = 0$ excluded). Inset: normalized standard deviation $\mathrm{Std}_{x \neq 0}(m_{2;x})/\mE_x(m_{2;x})$ as a function of $N$.
(c) Normalized standard deviation as a function of $N$ for $|\psi_m\rangle=|T\rangle^{\otimes N_T}\otimes|0\rangle^{\otimes(N-N_T)}$ before and after Clifford preconditioning.
(d) Normalized standard deviation as a function of step $N_C$ of Clifford preconditioning.
}
  \label{fig: random-magic}
\end{figure}

To compute the stabilizer R\'enyi entropy, after each FWHT, we accumulate the partial $2\alpha$-order moment
$m_{\alpha;x}:=\sum_{z\in\mF_2^N}|\langle\psi|P_{x,z}|\psi\rangle|^{2\alpha}/d^{\alpha}$, and hence
$M_\alpha(|\psi\rangle)=\frac{1}{1-\alpha}\log_2\!\big(\sum_x m_{\alpha;x}\big)-N$.
For the stabilizer nullity, it suffices to count the number of Pauli strings with $|\langle\psi|P_{x,z}|\psi\rangle|\approx 1$
(within a numerical tolerance, e.g.\ $10^{-7}$). %
A complete pseudocode is given in Algorithm~\ref{alg:FWHT}.
We benchmark the FWHT algorithm against brute-force Pauli enumeration on highly entangled random magic states $\mC \ket{\psi_m}$ ($\mC$ generated by $2N$ layers of Clifford circuit; see definition later). 
Both approaches agree with the analytic values up to numerical round-off, while the FWHT method is markedly faster, consistent with the expected scalings $\mO(N2^{2N})$ versus $\mO(2^{3N})$ [Fig.~\ref{fig: random-magic}\,(a)].

{\it Monte Carlo sampling with FWHT.---} %
We can approximate $M_{\alpha}$ by Monte Carlo (MC) sampling %
We always include $x=0$ and sample the remaining $\mN-1$ values uniformly from $\mF_2^N\setminus{0}$~\cite{supplemental}.
For each sampled $x$, we compute
$\{\langle\psi|P_{x,z}|\psi\rangle\}_{z\in\mF_2^N}$ via one FWHT and
accumulate the partial moment $m_{\alpha;x}$.
This yields the unbiased estimator
\begin{equation}
  \hat{S}:=
m_{\alpha;0}+\frac{d-1}{\mN-1}\sum_{x\in\mathrm{MC};\,x\neq 0}m_{\alpha;x}
\end{equation}
for $S:=\sum_{x\in\mF_2^N} m_{\alpha;x}$.
Notably, this procedure produces $\mN 2^N$ Pauli correlators at total cost $\mO(\mN N 2^N)$, i.e., an average cost $\mO(N)$ per sampled Pauli string, compared to $\mO(2^N)$ for direct evaluation of a single $\langle\psi|P|\psi\rangle$.

The MC efficiency is determined by the fluctuations
of $m_{\alpha;x}$.
For $\mN \ll d$, the relative standard deviation obeys $ \sigma_{\hat{S}}/{S} \approx \mN^{-1/2}\mathrm{Std}_{x \neq 0}(m_{\alpha;x})/\mE_x[m_{\alpha;x}]$, 
where $\mE_x[\cdot]$ is the mean over all $x\in\mF_2^N$ and $\mathrm{Std}_{x\neq 0}(\cdot)$ is the standard deviation over $x\in\mF_2^N\setminus\{0\}$.
As $M_\alpha=\frac{1}{1-\alpha}\log_2 S - N$, we have
\begin{equation}
\sigma_{\hat{M}_{\alpha}}
\approx
\frac{1}{|1-\alpha| \ln (2)\sqrt{\mN}}\frac{\mathrm{Std}_{x \neq 0}(m_{\alpha;x})}{\mE_x[m_{\alpha;x}]}.
\label{eq:stdM}
\end{equation}
From Eq.~\eqref{eq:stdM}, a target accuracy $\epsilon$ requires $\mN \sim \epsilon^{-2} (\mathrm{Std}_{x \neq 0}(m_{\alpha;x})/\mE_x[m_{\alpha;x}])^2$. In practice, one can estimate this ratio from a small pilot sample and then set $\mN$ accordingly.

\begin{figure}[tbhp]    
  \centering
  \includegraphics[width=0.48\textwidth]{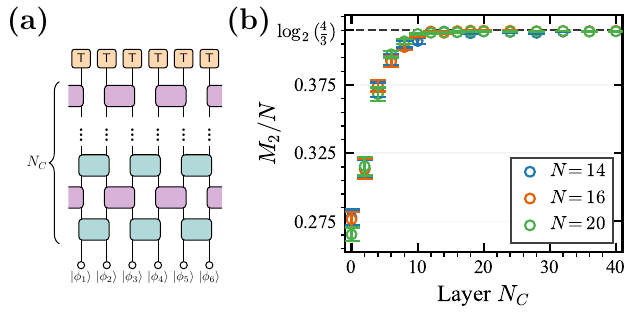}
  \caption{(a) Quantum circuit beginning with an $N$-qubit random Clifford product state, followed by $N_C$ layers of random two-qubit Clifford gates, and then $T$ gates on all qubits. (b) Stabilizer R\'enyi entropy density $M_2/N$ of the output states in panel~(a) as a function of $N_C$ for different system sizes $N$. The data are averaged over $80$ random instances. 
  For $N=14$ and $16$, we use the exact FWHT algorithm; for $N=20$, the number of Monte Carlo samples is $\mN=2 \times 10^4$.}
  \label{fig: Clifford T}
\end{figure}

We first consider Haar-random pure states as an example and focus on $\alpha=2$.
Numerically, we find that $m_{2;x}$ is well described by a Gaussian distribution and that
$\mathrm{Std}_{x \neq 0}(m_{2;x})/\mE_x[m_{2;x}]$ decreases exponentially with respect to $N$ [Fig.~\ref{fig: random-magic}\,(b); see also Ref.~\cite{szombathy2025a} for related concentration behavior], so that only a small $\mN$ is needed to accurately estimate $M_2$.
For example, with $\mN=10$ samples at $N=24$ we obtain $M_2=21.9999(2)$, in excellent agreement with the exact Haar value
$M_2^{\rm Haar}=\log_2(2^N+3)-2\approx 22$~\cite{leone2022,turkeshi2025a}.

\begin{figure*}[thbp]
  \centering
  \includegraphics[width=0.98\textwidth]{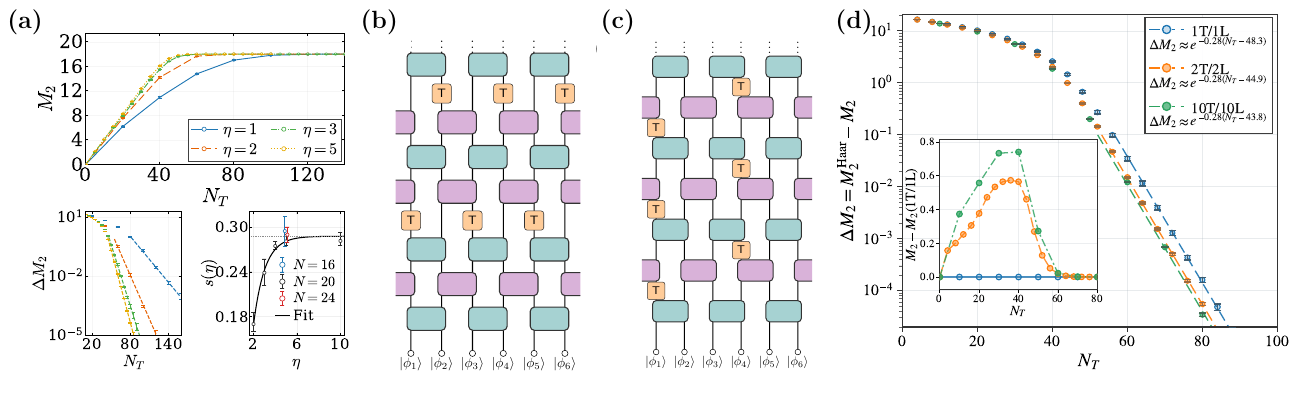}
\caption{
(a)~Top: ensemble-averaged $M_2$ versus the total number of $T$ gates $N_T$ for different scrambling ratios $\eta$ [Eq.~\eqref{eq:eta}] and $N = 20$.
Bottom left: the gap $\Delta M_2 = M_2^{\rm Haar} - \mE[M_2]$ on a logarithmic scale, showing exponential decay $\Delta M_2 \propto e^{-s(\eta)\, N_T}$.
Bottom right: decay rate $s(\eta)$ extracted from linear fits; error bars are obtained by bootstrapping~\cite{press2007}. 
The solid curve is a fit to $s(\eta) = s_\infty - \alpha \,e^{-\beta \eta}$ with $s_{\infty}$ fixed at $\ln(4/3)$ and fitted $\alpha \approx 0.79$, $\beta \approx 0.95$.
Additional points at $N = 16$, $20$, and $24$ for $\eta = 5$ show weak finite-size dependence.
(b,\,c)~Two $T$-injection schedules, illustrated for $N=6$ and $\eta = 3$: (b)~bursty ``3T/3L'' schedule ($3$ parallel $T$ gates every $3$ Clifford layers); (c)~uniform ``1T/1L'' schedule.
(d)~$\Delta M_2$ versus the total number of $T$ gates $N_T$ for three schedules (``1T/1L'', ``2T/2L'', ``10T/10L'') at fixed $\eta = 10$ and $N = 20$. All schedules share the same slope $s \approx 0.28$ but differ in the intercept $b$. Inset: difference $M_2 - M_2(\text{1T/1L})$, showing the advantage of burstier schedules.
MC sample size $\mN=2\times10^4$ and $80$ circuit realizations per data point throughout.
}
  \label{fig: Clifford T2}
\end{figure*}

{\it Clifford preconditioning for MC sampling.---}
For structured states, naive MC sampling can require exponentially many samples.
Consider the product magic state $|\psi_m\rangle=|T\rangle^{\otimes N_T}\otimes |0\rangle^{\otimes (N-N_T)}$ with
$|T\rangle=\frac{1}{\sqrt2}(|0\rangle+e^{i\pi/4}|1\rangle)$ and $N_T=\lfloor N/2\rfloor$,
which has $M_2=N_T\log_2(4/3)$, much smaller than $M_2^{\rm Haar}\approx N-2$~\cite{leone2022,turkeshi2025a}.
We analytically find $\mathrm{Std}_{x\neq 0}(m_{2;x})/\mE_x[m_{2;x}] \propto (20/9)^{N/4}$~\cite{supplemental} [also see Fig.~\ref{fig: random-magic}\,(c)], 
so the required $\mN$ grows as $(20/9)^{N/2}$ for fixed target accuracy.
This large standard deviation arises because different $x$-families have distinct support profiles (e.g., the $x=0$ family contains only $I/Z$ strings while the $x=1\cdots 1$ family contains only full-support strings), and in structured states, lower-support Pauli strings carry systematically larger weight.

A practical remedy is Clifford preconditioning. 
We apply a random brick-wall Clifford circuit $\mC$~\cite{corcoles2013} of depth $N_C$ before sampling.
Sampling on $\mC|\psi\rangle$ is equivalent to sampling the conjugated family $\{\mC^{-1}P_{x,z}\mC\}_{z}$ on the original state.
Also, $M_\alpha(\mC|\psi\rangle)=M_\alpha(|\psi\rangle)$.
For $N_C = \mO(N)$, the conjugated Pauli strings are no longer tied to their original support pattern~\cite{nahum2017b,nahum2018a}, giving each family a more mixed support profile and hence reducing $\mathrm{Std}_{x\neq 0}(m_{2;x})$.
Numerically, we find that $N_C \simeq 2N$ suffices to reduce $\mathrm{Std}_{x\neq 0}(m_{2;x})/\mE_x[m_{2;x}]$ to order unity, with no visible growth over $10 \leq N \leq 22$ [Figs.~\ref{fig: random-magic}\,(c) and (d)].
A moderate $\mN \sim 10^4$ then yields accuracy $\sim 10^{-2}$.
We have tested additional structured states, including output states of shallow Haar circuits and the ground state of the mixed-field Ising model, and find similar behavior after preconditioning~\cite{supplemental}.
This is in sharp contrast to the exponential scaling $\mN \propto (20/9)^{N/2}$ before preconditioning, suggesting that after preconditioning, $\mN$ for a fixed target accuracy does not need to scale with $N$ over the accessible system sizes, enabling more scalable estimation of $M_2$.

The overhead of preconditioning is modest. 
Each Clifford layer costs $\mO(N 2^N)$, comparable to one MC sample, so $N_C = 2N$ layers cost $\mO(N^2 2^N)$, equivalent to roughly $2N$ samples.
Since reliably estimating $\mathrm{Std}_{x\neq 0}(m_{2;x})$ itself requires a comparable number of samples, $N_C = 2N$ is a reasonable default choice when little prior knowledge about the state is available.
We note that the previously used Metropolis-Hastings sampling of individual Pauli strings~\cite{turkeshi2025a,metropolis1953,hastings1970} can become inefficient for states with low nonstabilizerness but high entanglement, such as $\mC|\psi_m\rangle$ with $N_C = 2N$ layers of Clifford scrambling; see Supplemental Material~\cite{supplemental} for a detailed comparison.

{\it $T$-gate injection and Clifford scrambling.---}
We study the injection of $T=\begin{pmatrix}1&0\\0&e^{i\pi/4}\end{pmatrix}$ gates into random Clifford circuits~\cite{zhou2020c,bejan2024,szombathy2025b,fux2024a,zhang2025g,ivaki2025,bejan2025}. 
We parameterize the interplay between scrambling and injection by the Clifford scrambling ratio
\begin{equation}
\eta := \frac{\text{number of two-qubit Clifford gates per cycle}}{\text{number of $T$ gates per cycle}} \, .
\label{eq:eta}
\end{equation}
As a first probe, we consider a single-shot protocol [Fig.~\ref{fig: Clifford T}\,(a)]:
starting from a random product stabilizer state $\bigotimes_{i=1}^N|\phi_i\rangle$ (each $|\phi_i\rangle$ drawn uniformly from the six single-qubit Clifford states), we apply $N_C$ layers of random brick-wall Clifford gates ($\eta = N_C/2$), then inject $T$ on every qubit.
The ensemble-averaged density $\mE[M_2/N]$ increases with $\eta$ and, for $N \ge 14$, collapses onto a single curve [Fig.~\ref{fig: Clifford T}\,(b)], with analytical limiting values $\mE[M_2/N] = 2/3 \log_2(4/3)$ at $\eta=0$ and $\log_2(4/3)$ as $\eta\to\infty$~\cite{supplemental}.
The saturated value $\log_2(4/3)$ per $T$ gate coincides with the nonstabilizing power of a single $T$ gate dressed by an independent random Clifford unitary~\cite{leone2024}, and with the per-$T$-gate contribution found in the dilute limit of $T$-doped circuits~\cite{szombathy2025b}.
Here, however, all $N$ $T$ gates are injected simultaneously after a single Clifford block, and saturation is reached already at $\eta \gtrsim 5$ (i.e., $N_C \gtrsim 10$ brick-wall layers), with no visible system-size dependence.

Even at saturation, $\mE[M_2/N] \approx \log_2(4/3) \approx 0.415$ is far below the Haar-random value (${\approx}\,1$ at large $N$)~\cite{turkeshi2025a}.
To approach it, we repeat the cycle ($N_C$ Clifford layers + $N$ parallel $T$ gates, $\eta = N_C/2$) and study how $\mE[M_2]$ grows with the total number of $T$ gates $N_T$ [Fig.~\ref{fig: Clifford T2}\,(a)].
Larger $\eta$ leads to faster growth at all stages, and the gap $\Delta M_2 := M_2^{\rm Haar} - \mE[M_2]$ decays exponentially at large $N_T$:
\begin{equation}
\ln \Delta M_2 \approx -s(\eta)\, N_T + b.
\label{eq:DeltaM2}
\end{equation}
The decay rate $s(\eta)$ increases with $\eta$ and is well described by $s(\eta)\approx s_\infty  - 0.79\,e^{-0.95\eta}$ [Fig.~\ref{fig: Clifford T2}\,(a)], where the asymptote $s_\infty = \ln(4/3)$ is obtained by Clifford-group averaging~\cite{supplemental}.
Already at $\eta \gtrsim 5$, $s(\eta)$ is close to $s_\infty$, and different system sizes ($N = 16$, $20$, $24$) yield nearly the same values, suggesting that this saturation is not a finite-size artifact.
The saturation also echoes the single-shot result, reinforcing the picture that modest Clifford scrambling suffices to fully realize the nonstabilizerness power (in terms of $M_2$) of $T$ gates.

We next fix $\eta$ and study the effect of the temporal distribution of $T$ gates on magic injection.
Concretely, each cycle consists of $k$ Clifford layers followed by $k$ parallel $T$ gates, and the cycle is repeated; the target qubits are approximately equally spaced within each injection layer and staggered between successive cycles [Fig.~\ref{fig: Clifford T2}\,(b,c)].
The label ``$k$T/$k$L'' denotes this schedule (e.g., ``10T/10L'' means 10 parallel $T$ gates after 10 Clifford layers per cycle).
The extreme $k=1$ corresponds to the most uniform injection, while larger $k$ concentrates $T$ gates into fewer, burstier rounds; in our numerics $\eta = N/2 = 10$, independent of $k$.
We find that all schedules share nearly the same decay rate $s$ for $\Delta M_2 = M_2^{\rm Haar} - \mE[M_2]$ [Fig.~\ref{fig: Clifford T2}\,(d)], indicating that $s$ is governed by $\eta$, while different temporal distributions affect only the intercept $b$ [Eq.~\eqref{eq:DeltaM2}; also see simulation for other $\eta$~\cite{supplemental}].
Burstier schedules achieve systematically higher $\mE[M_2]$ at the same $T$ budget, and quantitatively, the most concentrated schedule requires ${\sim}\,4$ fewer $T$ gates than the most uniform one to reach the same $M_2$.

{\it Summary and outlook.---}
We introduce an FWHT-based framework for evaluating stabilizer R\'enyi entropies and stabilizer nullity of generic $N$-qubit wavefunctions.
The average cost per sampled Pauli string is reduced from $\mO(2^N)$ to $\mO(N)$. 
The method operates directly on full wavefunctions, making it applicable to highly entangled states.
Combined with an MC scheme with Clifford preconditioning, the required sample number $\mN$ for a given target accuracy shows no visible growth with $N$ in all our benchmarks, enabling estimation of $M_2$ in even larger systems.

Applying the framework to $T$-doped random Clifford circuits, we identify the scrambling ratio $\eta$ as the key parameter governing magic growth.
Each $T$ gate is close to its full nonstabilizing power (in terms of $M_2$) with only modest Clifford scrambling ($\eta \gtrsim 5$), and this threshold shows no visible system-size dependence.
At fixed $\eta$, the temporal distribution of $T$ gates affects only the overall offset, not the growth rate, with burstier schedules being more resource-efficient.
Extending the analysis to two-qubit Haar random gates, we find that they can generate nonstabilizerness more efficiently at the same gate count (End Matter), though at higher experimental cost~\cite{chen2025a}.

More broadly, it would be interesting to apply the framework to nonequilibrium many-body problems, for example, to track how conserved quantities~\cite{rakovszky2018,rakovszky2019,zhou2020b,tirrito2025a} and dynamical constraints~\cite{lan2018} influence the growth of nonstabilizerness and its interplay with entanglement~\cite{viscardi2026,gu2025,tirrito2024}.
It would also be interesting to apply the framework to other Pauli-based diagnostics such as Bell magic~\cite{haug2023c}.
On the methodological side, a direct extension is to subsystem (mixed-state) stabilizer R\'enyi entropies via a partial FWHT scheme (Supplemental Material~\cite{supplemental}).
Further acceleration may be possible using sparse-FWHT methods~\cite{scheibler2015} when the Pauli spectrum is sparse, or by combining partial FWHT with sampling over both $x$ and $z$ sectors.
Another interesting direction is to explore FWHT-based classical analogues of Bell Difference Sampling, which has proved useful for inferring stabilizer structure~\cite{gross2021,grewal2024,grewal2025}.

{\it Note Added.---}
During the final stage of this manuscript, we became aware of a related and independent work~\cite{huang2025b}, which also uses FWHT to compute exact values of stabilizer R\'enyi entropies, but does not develop MC schemes and focuses on different physics applications. 
After submission, we also became aware of another independent work~\cite{sierant2026}, 
which uses the analogous fast Fourier transform on $\mathbb{Z}_d^N$ to compute 
qudit mana~\cite{veitch2014} for qutrits.

{\it Acknowledgement.---}
Z.X. thanks David Huse, Sarang Gopalakrishnan, and Shuo Liu for helpful discussions.
Z.X. is pleased to acknowledge that the work reported in this paper was substantially performed using the Princeton Research Computing resources at Princeton University.
Z.X. is supported by the Princeton Quantum Initiative Fellowship.
S.R. is supported by a Simons Investigator
Grant from the Simons Foundation (Award No.\ 566116).
This work is supported by the Gordon and Betty Moore Foundation EPiQS initiative, Grant GBMF8685.01.

\bibliography{ref_v2}

\onecolumngrid
\section*{End matter}
\twocolumngrid

\subsection{Pseudocode for FWHT-based Pauli sampling}
\label{sec:endmatter-pseudocode}
In this section, we describe the complete pseudocode to compute the stabilizer R\'enyi entropy and nullity by FWHT.

\begin{algorithm}[H]
\caption{Fast Walsh--Hadamard Pauli sampling}\label{alg:FWHT}
\begin{flushleft}
\textbf{Input:} An $N$-qubit wave function $|\psi\rangle=\sum_{b\in\mF_2^N}\psi(b)\,|b\rangle$ and numerical tolerance $\epsilon$ (for stabilizer nullity).\\
\textbf{Output:} Stabilizer R\'enyi entropy $M_\alpha$ and nullity $\nu$.
\end{flushleft}
\begin{algorithmic}[1]
\State Initialize $m_\alpha\gets 0$, $\nu_{\rm cnt}\gets 0$.
\For{$x\in\mF_2^N$}
    \State Define $f_x(b)\gets \overline{\psi(b)}\,\psi(b\oplus x)$ for all $b\in\mF_2^N$.
    \State Compute the FWHT $F_x(z)\gets \sum_{b\in\mF_2^N}(-1)^{z\cdot b}\,f_x(b)$ for all $z\in\mF_2^N$.
    \State $\nu_x\gets \#\{z\in\mF_2^N:\big||F_x(z)|-1\big|<\epsilon\}$.
    \State $m_{\alpha;x}\gets \sum_{z\in\mF_2^N} |F_x(z)|^{2\alpha}/2^{N\alpha}$.
    \State $m_\alpha\gets m_\alpha+m_{\alpha;x}$; \quad $\nu_{\rm cnt}\gets \nu_{\rm cnt}+\nu_x$.
\EndFor
\State \Return $M_\alpha=\frac{1}{1-\alpha}\log_2(m_\alpha)-N$,\quad $\nu=N-\log_2(\nu_{\rm cnt})$.
\end{algorithmic}
\end{algorithm}

\subsection{Doping with two-qubit Haar-random gates}
\label{sec:endmatter-twoqubit}

\begin{figure}[tbhp]
\centering
\includegraphics[width=0.48\textwidth]{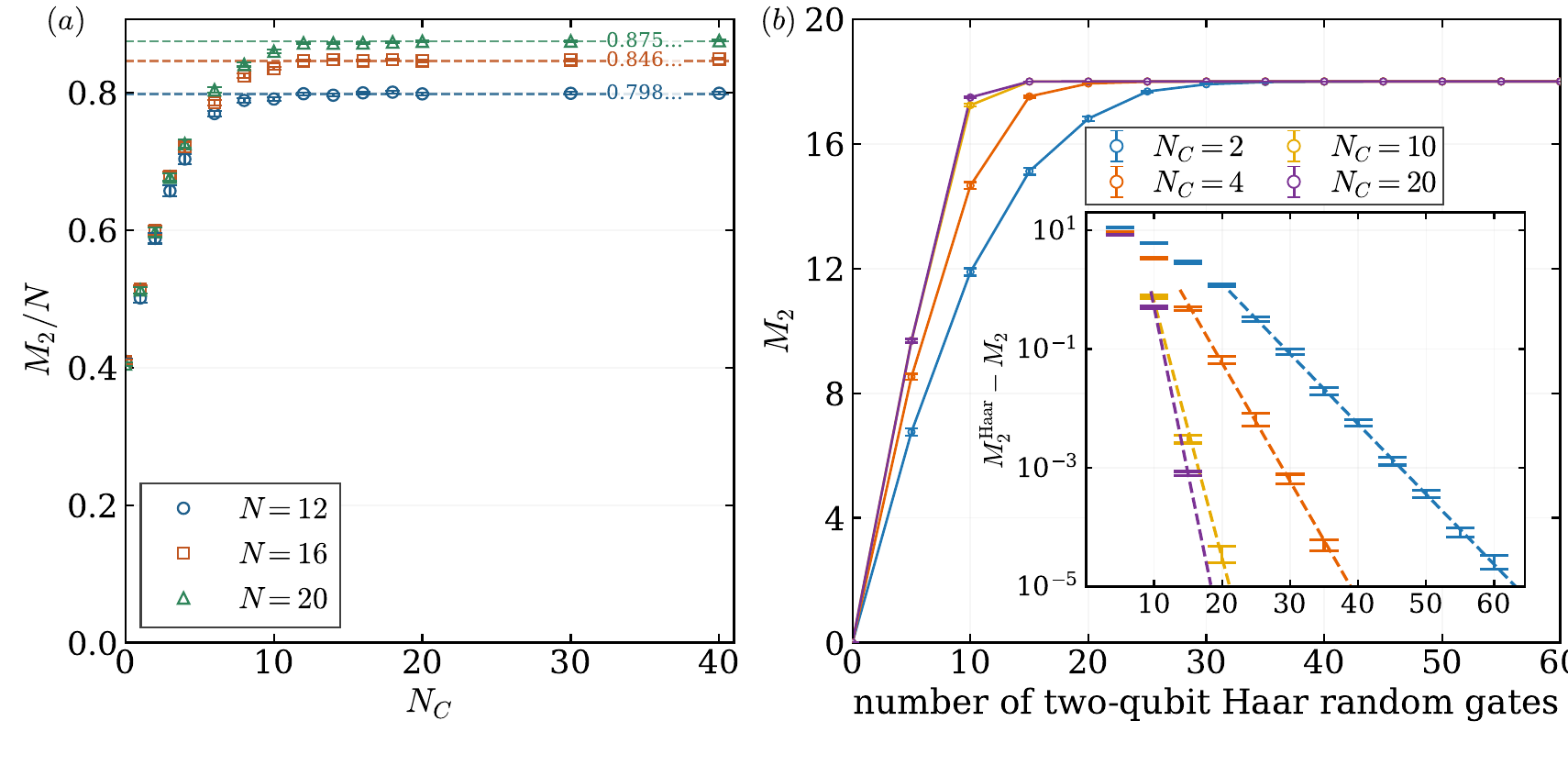}
\caption{
Nonstabilizerness generated by two-qubit Haar-random gates.
(a)~$M_2/N$ of the output state produced by $N_C$ layers of random two-qubit Clifford gates followed by a single layer of $N/2$ independent Haar-random two-qubit gates on neighboring pairs.
Dashed lines: analytical values in the large-$N_C$ limit~\cite{supplemental}.
(b)~$M_2$ versus the total number of two-qubit Haar gates, obtained by repeating the cycle of $N_C$ Clifford layers and one Haar-gate layer.
For $N = 12$, $16$, $M_2$ is computed exactly; for $N = 20$, MC sample size $\mN = 2 \times 10^4$.
Each data point is averaged over $80$ circuit realizations.}
\label{fig:M2Haar}
\end{figure}

We extend the analysis of the main text by replacing single-qubit $T$ gates with two-qubit Haar-random gates.
We apply $N/2$ independent Haar-random two-qubit gates on neighboring pairs, interleaved with $N_C$ layers of random brick-wall Clifford gates.

As in the $T$-gate case, $\mE[M_2/N]$ increases with $N_C$ and saturates at $N_C \gtrsim 15$ [Fig.~\ref{fig:M2Haar}(a)].
Even without Clifford scrambling ($N_C = 0$), we analytically find $\mE[M_2] = N \log_2(7/4)/2 \approx 0.403\,N$~\cite{supplemental}, comparable to the saturated value of the $T$-gate protocol ($\mE[M_2] \approx N \log_2(4/3) \approx 0.415\,N$; consistent with numerics), despite using only $N/2$ two-qubit gates compared to $N$ single-qubit $T$ gates.
At large $N_C$, the saturated value $\mE[M_2]/N \approx 0.875$ far exceeds the $T$-gate case~\cite{supplemental}.

The gap $\Delta M_2 = M_2^{\rm Haar} - \mE[M_2]$ again decays exponentially with the total number of injected gates [Fig.~\ref{fig:M2Haar}(b)], but at a much faster rate than for $T$ gates.
For example, at $N = 20$ and $N_C = 4$, reaching $\Delta M_2 = 10^{-5}$ requires around $40$ two-qubit Haar gates versus around $120$ $T$ gates.
These results show that two-qubit Haar random gates are significantly more efficient for magic generation, though at higher experimental cost~\cite{chen2025a}.
\end{document}


\setcounter{equation}{0}
\setcounter{section}{0}
\setcounter{figure}{0}
\setcounter{table}{0}
\setcounter{page}{1}
\renewcommand{\theequation}{S\arabic{equation}}
\renewcommand{\thesection}{\Roman{section}}

\renewcommand{\thefigure}{S\arabic{figure}}
\renewcommand{\thetable}{\arabic{table}}
\renewcommand{\tablename}{Supplementary Table}

\renewcommand{\bibnumfmt}[1]{[S#1]}
\renewcommand{\citenumfont}[1]{#1}

\title{Supplemental Material for ``Exponentially Accelerated Sampling of Pauli Strings for Nonstabilizerness''}
    
\author{Zhenyu Xiao}
\email{zyxiao@princeton.edu}
\affiliation{Princeton Quantum Initiative, Princeton University, Princeton, New Jersey 08544, USA}

\author{Shinsei Ryu}
\affiliation{Department of Physics, Princeton University, Princeton, New Jersey 08544, USA}

\date{\today}
	
\maketitle

\tableofcontents

\section{Monte Carlo sampling} %
We approximate $M_{\alpha}$ by Monte Carlo (MC) sampling over $x\in\mF_2^N$. %
As described in the main text, we always include $x=0$ deterministically and sample the remaining $\mN-1$ values of $x$ uniformly at random from $\mF_2^N\setminus\{0\}$.
For each sampled $x$, we compute $\{\langle\psi|P_{x,z}|\psi\rangle\}_{z\in\mF_2^N}$ via one FWHT and accumulate the partial moment
$m_{\alpha;x}:=\sum_{z\in\mF_2^N}|\langle\psi|P_{x,z}|\psi\rangle|^{2\alpha}/d^{\alpha}$.

We treat $x=0$ separately because the $x=0$ family is atypical. 
It contains the identity operator $I$, for which $\langle\psi|I|\psi\rangle=1$.
For $\alpha=2$, this contribution can be comparable to the aggregate contribution from all $P\neq I$ in Haar-random states~\cite{leone2022}, and including $x=0$ in the random pool increases sampling fluctuations.
By always including $x=0$ and sampling only from $x\neq 0$, we obtain the unbiased estimator
\begin{equation}
\hat S:= m_{\alpha;0}+\frac{d-1}{\mN-1}\sum_{x\in\mathrm{MC};\,x\neq0} m_{\alpha;x}
\end{equation}
for $S:=\sum_{x\in\mF_2^N} m_{\alpha;x}$, and hence $\hat M_{\alpha}=\frac{1}{1-\alpha}\log_2\hat S-N$.
For $\mN\ll d$, the relative standard deviation of $\hat{S}$ satisfies $\sigma_{\hat{S}}/S \approx \mN^{-1/2}\,\mathrm{Std}_{x\neq 0}(m_{\alpha;x})/\mE_x[m_{\alpha;x}]$,
where $\mE_x[\cdot]$ is the mean over all $x\in\mF_2^N$ and $\mathrm{Std}_{x\neq 0}(\cdot)$ is the standard deviation over $x\in\mF_2^N\setminus\{0\}$.

\section{Clifford preconditioning: analytics and benchmarks}
\label{sec:sm-precond-bench}

\subsection{Analytic fluctuation ratio for the product magic state}
\label{sec:sm-psi-m-exact}

We derive the fluctuation ratio $r_2=\mathrm{Std}_{x\neq 0}(m_{2;x})/\mE_x[m_{2;x}]$ for the unscrambled product magic state
$\ket{\psi_m}=\ket{T}^{\otimes N/2}\otimes \ket{0}^{\otimes N/2}$
with $\ket{T}=(\ket{0}+e^{i\pi/4}\ket{1})/\sqrt{2}$, assuming even $N$ for simplicity (the large-$N$ scaling is unchanged for odd $N$).

Since $\ket{\psi_m}$ is a product state, $m_{2;x}=4^{-N}\sum_{z}|\bra{\psi_m}P_{x,z}\ket{\psi_m}|^4$ factorizes over qubits.
Writing $x=(x_A,x_B)$ with $x_A,x_B\in\mF_2^{N/2}$ for the $\ket{T}$ and $\ket{0}$ sectors respectively, a direct evaluation of the single-qubit Pauli expectation values gives the local factors:
for each $\ket{T}$ qubit,
$|\bra{T}I\ket{T}|^4+|\bra{T}Z\ket{T}|^4=1$ and
$|\bra{T}X\ket{T}|^4+|\bra{T}Y\ket{T}|^4=1/2$;
for each $\ket{0}$ qubit,
$|\bra{0}I\ket{0}|^4+|\bra{0}Z\ket{0}|^4=2$ and
$|\bra{0}X\ket{0}|^4+|\bra{0}Y\ket{0}|^4=0$.
Hence $m_{2;x}=0$ whenever $x_B\neq 0$; if $x_B=0$ and $x_A$ has Hamming weight $k$,
\begin{equation}
m_{2;x}
=
2^{-3N/2-k}.
\label{eq:sm-psi-m-m2x}
\end{equation}

Summing over the $\binom{N/2}{k}$ such labels gives
\begin{equation}
\mE_x[m_{2;x}]
=
\frac{1}{4^{N/2}}
\sum_{k=0}^{N/2}
\binom{N/2}{k}2^{-3N/2-k}
=
\left(\frac{3}{64}\right)^{N/2}.
\label{eq:sm-psi-m-mean-all}
\end{equation}
The first two moments over $x\neq 0$ are
$\mE_{x\neq 0}[m_{2;x}]
=
2^{-3N/2}
\bigl[(3/2)^{N/2}-1\bigr]/(4^{N/2}-1)$
and
$\mE_{x\neq 0}[m_{2;x}^2]
=
2^{-3N}
\bigl[(5/4)^{N/2}-1\bigr]/(4^{N/2}-1)$,
yielding the exact ratio
\begin{equation}
r_2
=
\left(\frac{8}{3}\right)^{N/2}
\sqrt{
\frac{(5/4)^{N/2}-1}{4^{N/2}-1}
-
\frac{\left[(3/2)^{N/2}-1\right]^2}{(4^{N/2}-1)^2}
}.
\label{eq:sm-psi-m-ratio-exact}
\end{equation}
At large $N$, the first term under the square root dominates, giving
\begin{equation}
r_2
\sim
\left(\frac{20}{9}\right)^{N/4}.
\label{eq:sm-psi-m-ratio-asym}
\end{equation}
This is the exponential scaling quoted in the main text before Clifford preconditioning.

\subsection{Brick-wall circuits and Ising ground states}
The main text demonstrates Clifford preconditioning on the product magic state $|\psi_m\rangle$.
In this section, we test the MC scheme on two additional classes of structured states:
\begin{enumerate}[label=(\roman*)]
\item Output states of random brick-wall circuits of depth $d_H=1,2,4,6$, prepared from the input state $|0\rangle^{\otimes N}$.
Each brick-wall layer consists of independent Haar-random two-qubit gates on nearest-neighbor pairs; even and odd layers count separately.
\item Ground states of the open-chain mixed-field Ising model (MFIM)
\begin{equation}
H
=
g \sum_{i=1}^{N} \sigma_i^x
+ h \sum_{i=2}^{N-1} \sigma_i^z
+ (h-J)(\sigma_1^z + \sigma_N^z)
+ J \sum_{i=1}^{N-1} \sigma_i^z \sigma_{i+1}^z ,
\end{equation}
with $g = (\sqrt{5}+5)/8$, $h = (\sqrt{5}+1)/4$, and $J = 1$.
\end{enumerate}
Throughout this section, we monitor the normalized fluctuation
\begin{equation}
r_2
:=
\frac{\mathrm{Std}_{x\neq 0}(m_{2;x})}{\mE_x[m_{2;x}]},
\end{equation}
which controls the required MC sample number through $\mN \sim r_2^2/\varepsilon^2$ for a target accuracy $\varepsilon$ (see the main text for details).

\begin{figure}[t]
\centering
\includegraphics[width=\textwidth]{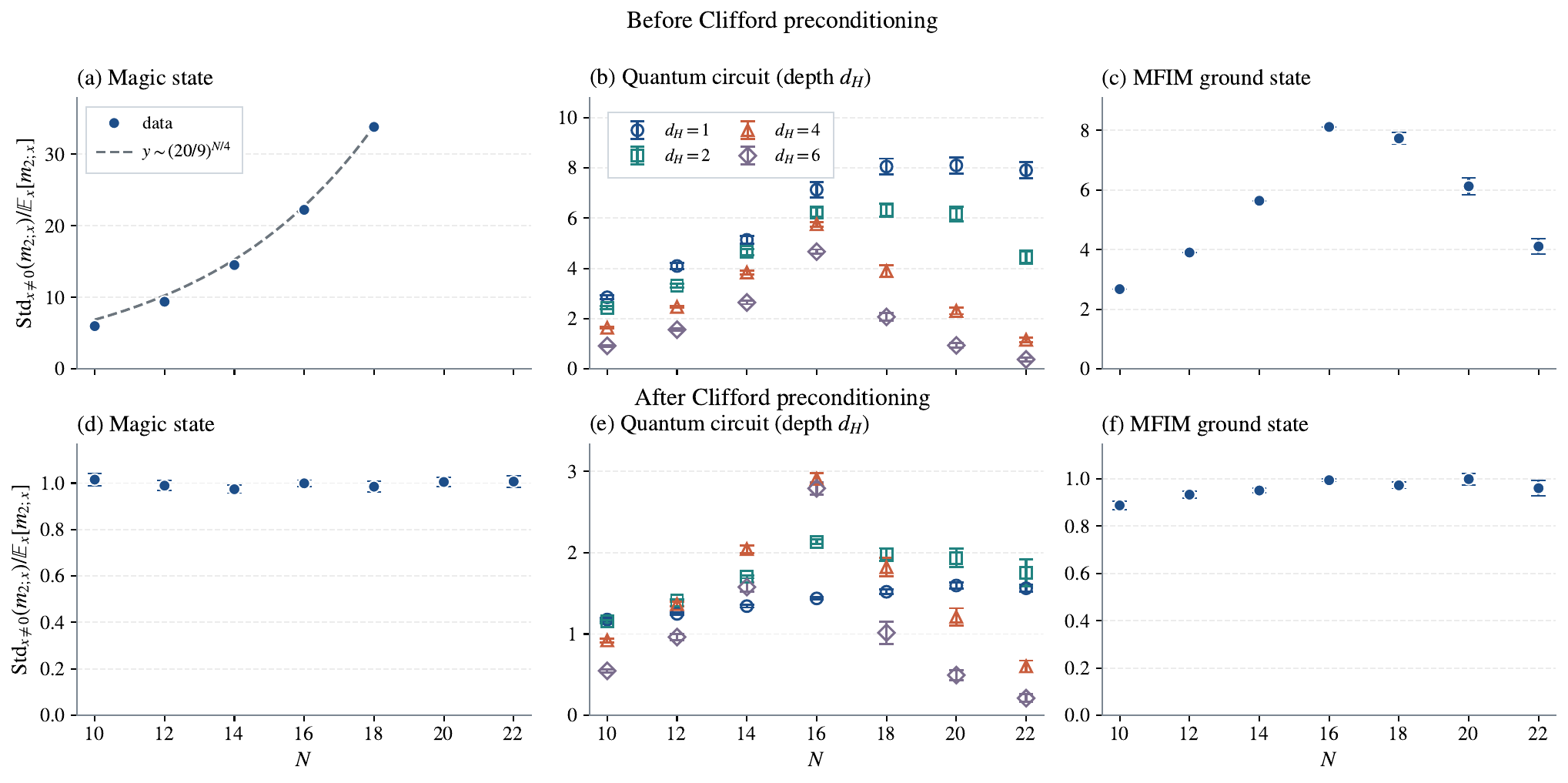}
\caption{
The normalized fluctuation $r_2=\mathrm{Std}_{x\neq 0}(m_{2;x})/\mE_x[m_{2;x}]$
before (top row) and after (bottom row) Clifford preconditioning with depth $N_C=2N$.
(a,\,d)~Product magic state
$|\psi_m\rangle = |T\rangle^{\otimes N_T}\otimes|0\rangle^{\otimes(N-N_T)}$ with
$N_T=\lfloor N/2\rfloor$.
(b,\,e)~Output states of Haar-random brick-wall circuits of depth
$d_H=1,2,4,6$ applied to $|0\rangle^{\otimes N}$.
(c,\,f)~Ground states of the MFIM.
Each data point is averaged over $80$ random realizations; for $N\le 16$ the Pauli spectrum is computed exactly, while for $N>16$ it is estimated with $2\times 10^4$ MC samples.
}
\label{fig:sm-precond-overview}
\end{figure}

We first investigate the size dependence of $r_2$ before and after preconditioning.
Figure~\ref{fig:sm-precond-overview} compares $r_2$ as a function of system size $N$ before and after Clifford preconditioning with depth $N_C=2N$.
Panels (a) and (d) reproduce the product-magic-state benchmark and are included as a reference.

For the brick-wall circuit states [panels (b) and (e)], the unpreconditioned $r_2$ depends strongly on both $d_H$ and $N$, and is non-monotonic at small system sizes.
After preconditioning, $r_2$ is reduced to below $3$ in all cases, so that $\mN \sim 10^4$ suffices for an accuracy $\varepsilon \sim 0.03$.
Importantly, $r_2$ after preconditioning shows no clear increasing trend with $N$; for $N\ge 16$, it either stays flat or decays, suggesting that $\mN$ does not need to grow with $N$ to maintain a fixed accuracy.
Notably, for larger system sizes, $r_2$ decreases with $d_H$ when $d_H\ge 2$: the MC estimator becomes more efficient for deeper circuits.
This can be understood from the fact that larger $d_H$ leads to larger $M_2$, corresponding to a more uniform Pauli weight distribution and hence smaller $r_2$.
These observations are consistent with the expectation that more structured, lower-magic states provide a more demanding benchmark for the MC scheme.

For the MFIM ground state [panels (c) and (f)], the effect of preconditioning is even more pronounced: $r_2$ remains below $1$ for all system sizes $N=10,12,\dots,22$ after preconditioning, confirming that the MC estimation is also efficient for this physically relevant class of states.

\begin{figure}[t]
\centering
\includegraphics[width=\textwidth]{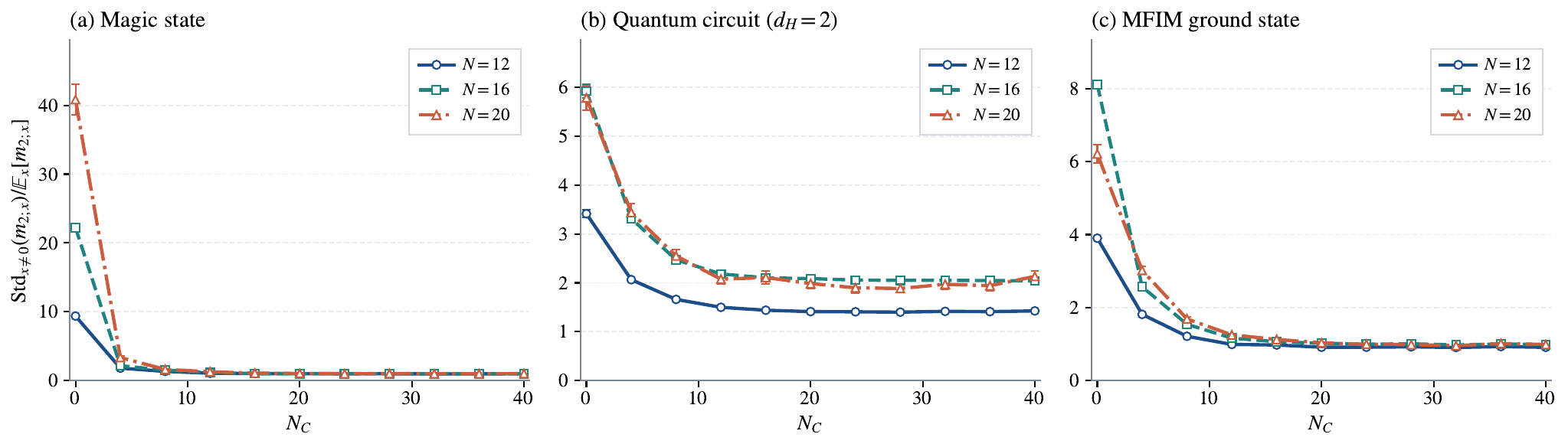}
\caption{
The normalized fluctuation $r_2$ as a function of the Clifford preconditioning depth $N_C$ for $N=12, 16, 20$.
(a)~Product magic state.
(b)~Output states of a Haar-random brick-wall circuit with depth $d_H=2$.
(c)~Ground states of the MFIM.
Each data point is averaged over $80$ random realizations; for $N\le 16$ the Pauli spectrum is computed exactly, while for $N>16$ it is estimated with $2\times 10^4$ MC samples.
}
\label{fig:sm-precond-traj}
\end{figure}

We next examine how $r_2$ depends on the preconditioning depth.
Figure~\ref{fig:sm-precond-traj} shows how $r_2$ decreases as $N_C$ increases, for three representative system sizes $N=12, 16, 20$.
In all three classes of states, the reduction is rapid: $r_2$ is already close to its saturated value by $N_C\approx 2N$, with no strong dependence on $N$.
This confirms that the choice $N_C=2N$ adopted in the main text is a practical default that is neither wasteful nor insufficient for the structured states considered here.

\section{Direct Pauli-string sampling with Metropolis--Hastings updates}

We describe the Metropolis--Hastings (MH) sampling of individual Pauli strings used for comparison in the main text.
Introducing the target distribution $\pi(P):=|\langle\psi|P|\psi\rangle|^2/d$, the sum $S=\sum_x m_{\alpha;x}=\frac{1}{d}\sum_P |\langle\psi|P|\psi\rangle|^{2\alpha}$ defined in the main text can be rewritten as
\begin{equation}
S=\frac{1}{d}\sum_P |\langle\psi|P|\psi\rangle|^{2\alpha}
=
\mE_{P\sim\pi}\!\left[|\langle\psi|P|\psi\rangle|^{2\alpha-2}\right].
\end{equation}
Accordingly, after collecting $\mN$ samples $P^{(1)},\dots,P^{(\mN)}$ from a Markov chain whose stationary distribution is $\pi$, one estimates $S$ by
\begin{equation}
\hat{S}_{\mathrm{MH}}=\frac{1}{\mN}\sum_{t=1}^{\mN} |\langle\psi|P^{(t)}|\psi\rangle|^{2\alpha-2},
\end{equation}
and then obtains $M_\alpha=\frac{1}{1-\alpha}\log_2 \hat{S}_{\mathrm{MH}}-N$.

Each Pauli string on $N$ qubits can be written as $P=\bigotimes_{j=1}^{N}\sigma_j$ with $\sigma_j\in\{I,X,Y,Z\}$.
At every step of the chain, a site $j$ is chosen uniformly at random, and the local operator $\sigma_j$ is replaced by one of the other three choices in $\{I,X,Y,Z\}$, also chosen uniformly at random, yielding a candidate string $\widetilde P$.
Because the proposal is symmetric (the probability of proposing $\widetilde P$ from $P$ equals that of proposing $P$ from $\widetilde P$), the Metropolis--Hastings acceptance probability reduces to
\begin{equation}
p_{\mathrm{acc}}
=
\min\!\left\{
1,\;\frac{\pi(\widetilde P)}{\pi(P)}
\right\}
=
\min\!\left\{
1,\;\frac{|\langle\psi|\widetilde P|\psi\rangle|^2}{|\langle\psi|P|\psi\rangle|^2}
\right\}.
\end{equation}
With probability $p_{\mathrm{acc}}$ the chain moves to $\widetilde P$; otherwise it stays at $P$.
Each proposed move requires evaluating one Pauli expectation value $\langle\psi|\widetilde P|\psi\rangle$ for the full $N$-qubit state vector, which costs $\mO(2^N)$ operations.
By contrast, in the MC+FWHT scheme the amortized cost per Pauli string is $\mO(N)$.

The chain is initialized by drawing $P^{(0)}$ uniformly at random and redrawing until $\pi(P^{(0)})$ exceeds a numerical threshold ($10^{-14}$ in practice).
The first $\mN_{\mathrm{burn}}$ steps (we use $\mN_{\mathrm{burn}}=10^4$) are discarded as burn-in before recording samples.
The acceptance ratio reported in the main text is the fraction of proposed updates that are accepted over the post-burn-in portion of the chain.
When the acceptance ratio is small, consecutive samples are highly correlated, and the effective number of independent samples $\mN_{\mathrm{eff}}$ can be much smaller than $\mN$.

\begin{figure}[t]
\centering
\includegraphics[width=0.38\textwidth]{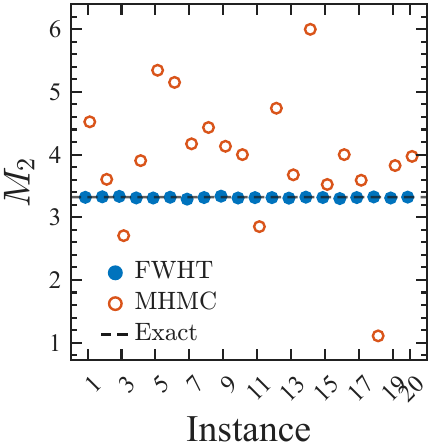}
\caption{
Fluctuations of $M_2$ estimates for highly entangled $\mC\ket{\psi_m}$ at $N=16$.
MC+FWHT samples the family $\{P_{x,z}\}_{z\in\mF_2^N}$ with $\mN=10^4$, while Metropolis--Hastings samples individual Pauli strings ($1.6\times 10^6$ samples).
The Metropolis--Hastings data uses $10\times$ more CPU time.
}
\label{fig:sm-mhmc}
\end{figure}

The combination of low acceptance and the $\mO(2^N)$ per-string evaluation cost makes MH sampling substantially less efficient than MC+FWHT for low-magic, highly entangled states.
As a concrete comparison, Figure~\ref{fig:sm-mhmc} shows independent estimates of $M_2$ for the state $\mC\ket{\psi_m}$ at $N=16$, where $\ket{\psi_m}=\ket{T}^{\otimes N_T}\otimes\ket{0}^{\otimes(N-N_T)}$ and $\mC$ is a random Clifford circuit of depth $N_C=2N$.
The MH acceptance ratio for this state is only $\approx 0.03$.
Despite using $10\times$ more CPU time, the MH estimates exhibit substantially larger fluctuations than the MC+FWHT results.

\section{Nonstabilizerness power of extensive $T$ gates}
\label{sec: T gates}

\subsection{Single-shot protocol}
\label{sec: T gates single}
We start from a random product stabilizer state
$\ket{\Phi}=\bigotimes_{i=1}^N\ket{\phi_i}$, where each $\ket{\phi_i}$ is drawn uniformly from the six single-qubit Clifford states (eigenstates of $X$, $Y$, and $Z$).
We then apply $N_C$ layers of random two-qubit Clifford gates in a brick-wall pattern, which defines a random Clifford circuit $\mC$ and prepares an entangled stabilizer state
$\ket{\psi(N_C)}=\mC\ket{\Phi}$.
Finally, we apply a $T$ gate to each qubit to obtain $T^{\otimes N}\ket{\psi(N_C)}$ and analyze the resulting extensive nonstabilizerness.

As a baseline, consider the no-scrambling limit $N_C=0$, in which $T^{\otimes N}\ket{\Phi}=\bigotimes_{i=1}^N T\ket{\phi_i}$ remains a product state.
Additivity of $M_2$ under tensor products then gives
\begin{equation}
M_2\!\left(\bigotimes_{i=1}^N T\ket{\phi_i}\right)
=
\sum_{i=1}^N M_2\!\left(T\ket{\phi_i}\right).
\end{equation}
For each site, $\ket{\phi_i}$ is a $Z$ eigenstate with probability $1/3$, in which case $T\ket{\phi_i}$ is a stabilizer state and $M_2(T\ket{\phi_i})=0$; otherwise $M_2(T\ket{\phi_i})=\log_2(4/3)$~\cite{leone2022}.
Averaging over the six initial Clifford states therefore yields
\begin{equation}
\mE\!\left[M_2/N\right]_{N_C=0}
=\frac{2}{3}\log_2(4/3).
\label{eq:M2_NC0}
\end{equation}
We then consider large $N_C$, where $\ket{\psi(N_C)}$ becomes entangled.

For analytical tractability, we work with the stabilizer linear entropy
\begin{equation}
M_{\rm lin}(\ket{\psi})
:=1-\frac{1}{d}\sum_{P}|c_P|^4,
\qquad
c_P=\langle\psi|P|\psi\rangle,
\qquad
d=2^N.
\end{equation}
For pure states, $M_{\rm lin}$ is related to the second-order stabilizer R\'enyi entropy by
$M_2=-\log_2(1-M_{\rm lin})$.
It is convenient to rewrite $M_{\rm lin}$ in replica form as
$M_{\rm lin}(\ket{\psi})=1-d\Tr\!\big[(\ket{\psi}\!\bra{\psi})^{\otimes4}Q\big]$,
where we introduce the operator
$Q:=\frac{1}{d^2}\sum_{P} P^{\otimes4}$.

We consider the circuit-averaged quantity
\begin{equation}
\mathbb{E}\!\left[M_{\rm lin}\!\left(T^{\otimes N}\ket{\psi(N_C)}\right)\right],
\end{equation}
where $\mathbb{E}[\cdot]$ denotes the average over the random brick-wall Clifford circuit.
For sufficiently large $N_C$, the induced distribution of the circuit $\mC$ is well approximated by the uniform distribution over the $N$-qubit Clifford group. We therefore replace $\mathbb{E}$ by the Clifford-group average $\mathbb{E}_{\mC}$.

Using $\ket{\psi(N_C)}=\mC\ket{\Phi}$ and the replica representation of $M_{\rm lin}$, we obtain
\begin{equation}
\mathbb{E}_{\mC}\!\left[M_{\rm lin}\!\left(T^{\otimes N}\ket{\psi(N_C)}\right)\right]
=1-d\,\mathbb{E}_{\mC}\,
\Tr\!\left[
Q\,
\big(T^{\otimes N}\mC\ket{\Phi}\!\bra{\Phi}\mC^\dagger T^{\dagger\otimes N}\big)^{\otimes4}
\right].
\label{eq:Em_Mlin_replica}
\end{equation}
For a stabilizer input $\ket{\Phi}$, we have~\cite{leone2021a}
\begin{equation}
\mathbb{E}_{\mC}\!\left[\big(\mC\ket{\Phi}\!\bra{\Phi}\mC^\dagger\big)^{\otimes4}\right]
= a\,\Pi_{\rm sym}+b\,Q\Pi_{\rm sym}.
\label{eq:clifford_twirl_4copy}
\end{equation}
Here, $\Pi_{\rm sym}:=\frac{1}{24}\sum_{\pi\in S_4} W_\pi$ is the projector onto the fully symmetric subspace of the $4$-replica Hilbert space,
with $W_\pi$ the permutation operator acting on basis states as
$W_\pi\ket{i_1 i_2 i_3 i_4}=\ket{i_{\pi^{-1}(1)}\, i_{\pi^{-1}(2)}\, i_{\pi^{-1}(3)}\, i_{\pi^{-1}(4)}}$.
The coefficients $a$ and $b$ are
\begin{equation}
a=\frac{24}{d(d+1)(d+2)(d+4)},
\qquad
b=\frac{6}{(d+1)(d+2)(d+4)}.
\label{eq:ab_values}
\end{equation}
Substituting Eq.~\eqref{eq:clifford_twirl_4copy} into Eq.~\eqref{eq:Em_Mlin_replica} and using the fact that $\Pi_{\rm sym}$ commutes with $T^{\otimes 4N}$, we find
\begin{equation}
\mathbb{E}_{\mC}\!\left[M_{\rm lin}\right]
=1-d\left[
a\,\Tr\!\left(Q\Pi_{\rm sym}\right)
+b\,\Tr\!\left(Q\,T^{\otimes 4N} Q\,T^{\dagger \otimes 4N}\,\Pi_{\rm sym}\right)
\right].
\label{eq:Em_Mlin_after_twirl}
\end{equation}

To evaluate the remaining trace, we decompose $Q$ as $ Q=q^{\otimes N}$ with $q=\frac{1}{4}\Big(I^{\otimes4}+X^{\otimes4}+Y^{\otimes4}+Z^{\otimes4}\Big)$,
similarly decompose permutation operators as $W_\pi=\big(W_\pi^{(1)}\big)^{\otimes N}$, where $W_\pi^{(1)}$ acts on the single-qubit $4$-replica space. Then, we have
\begin{equation}
\Tr\!\left(Q\,T^{\otimes 4N} Q\,T^{\dagger \otimes 4N}\,\Pi_{\rm sym}\right)
=\frac{1}{24}\sum_{\pi\in S_4}
\left[
\Tr\!\left(T^{\otimes4} q\,T^{\dagger\otimes4}\,q\,W_\pi^{(1)}\right)
\right]^N.
\end{equation}
After some algebra, Eq.~\eqref{eq:Em_Mlin_after_twirl} simplifies to
\begin{equation}
\mathbb{E}_{\mC}\! \left[M_{\rm lin}\!\left(T^{\otimes N}\mC\ket{\Phi}\right) \right]
=
1-\frac{4(d-1)}{d^2+3d-4}
-\frac{d(d-1)\,(3^N+3)}{(d+1)(d+2)\,(d^2+3d-4)} =
1-\left(3/4 \right)^{N}
-{4}/{2^{N}} + \mathcal{O}\left(\left( {3}/{8}\right)^N\right) \, .
\label{eq:final_result_Mlin_TCliff}
\end{equation}
If the fluctuation of $M_{\rm lin}$ is small, we can approximate $\mE_{\mC} [ M_2] \approx -\log_2\left(1- \mE_{\mC}[M_{\rm lin}] \right) = N \log_2(4/3) + \mathcal{O}((2/3)^{N})$, which is consistent with the numerical results of large $N_C$ in the main text.

\subsection{Multi-round protocol}
\label{sec: T gates multi}

We now extend the analysis to $m$ rounds of $T$-gate injection, each preceded by an independent Clifford-group average.
Let $\tau:=T^{\otimes N}$ and define the $m$-round unitary
\begin{equation}
U_m := \tau\,\mC_m \cdots \tau\,\mC_2\,\tau\,\mC_1,
\end{equation}
where $\mC_1,\dots,\mC_m$ are independent random variables drawn uniformly from the $N$-qubit Clifford group.
We take the initial state as $\ket{\Phi}=\bigotimes_{i=1}^N\ket{\phi_i}$ and define the averaged four-copy state
\begin{equation}
\overline{\rho}_m
:=
\mE_{\mC_1,\dots,\mC_m}
\left[
\left(
U_m \ket{\Phi}\!\bra{\Phi} U_m^\dagger
\right)^{\otimes 4}
\right],
\end{equation}
together with the scalar
\begin{equation}
x_m := d\,\Tr(Q\,\overline{\rho}_m) = 1 - \mE_{\mC_1,\dots,\mC_m}\!\left[M_{\rm lin}(U_m\ket{\Phi})\right].
\label{eq:xm_def}
\end{equation}

We derive a closed-form expression for $x_m$ by establishing a one-round scalar recursion.
Let $\mathcal{T}(X):=\mE_{\mC}[\mC^{\otimes 4} X\, \mC^{\dagger\otimes 4}]$ be the four-copy Clifford twirling channel.
For any permutation-symmetric four-copy operator $X$,
the output $\mathcal{T}(X)$ is again permutation-symmetric and, moreover,
Clifford-invariant.
For qubits, the permutation-symmetric Clifford-invariant subspace of the
four-copy space is two-dimensional, spanned by $\Pi_{\rm sym}$ and
$Q\Pi_{\rm sym}$~\cite{leone2021a,gross2021}
(the stabilizer-state average in Eq.~\eqref{eq:clifford_twirl_4copy} is a special case).
Therefore, for any such $X$ with $\Tr(X)=1$ and $x:=d\,\Tr(QX)$,
\begin{equation}
\mathcal{T}(X)
=
\alpha(x)\,\Pi_{\rm sym} + \beta(x)\,Q\Pi_{\rm sym},
\label{eq:twirl_general}
\end{equation}
where $\Pi_{\rm sym}$ and $Q$ are defined in Sec.~\ref{sec: T gates single}.
The coefficients are fixed by the two scalar constraints
$\Tr(\mathcal{T}(X))=1$ and $d\,\Tr(Q\,\mathcal{T}(X))=x$
(the latter follows from $Q$ being Clifford-invariant), namely
\begin{equation}
\begin{aligned}
\alpha(x)\,\Tr(\Pi_{\rm sym})
+ \beta(x)\,\Tr(Q\Pi_{\rm sym})
&= 1, \\
\left[\alpha(x)+\beta(x)\right]\,d\,\Tr(Q\Pi_{\rm sym})
&= x,
\end{aligned}
\end{equation}
where in the second line we used $Q^2=Q$.
Using $\Tr(\Pi_{\rm sym})=d(d+1)(d+2)(d+3)/24$ and
$\Tr(Q\Pi_{\rm sym})=(d+1)(d+2)/6$, we solve this $2\times 2$
system to obtain
\begin{equation}
\alpha(x) = \frac{24(d-x)}{d(d-1)(d+1)(d+2)(d+4)},
\qquad
\beta(x) = \frac{6[(d+3)x-4]}{(d-1)(d+1)(d+2)(d+4)}.
\label{eq:alpha_beta_x}
\end{equation}

Given $\overline{\rho}_m$, the next round applies a Clifford twirl followed by a $T$ layer:
$\overline{\rho}_{m+1} = \tau^{\otimes 4}\,\mathcal{T}(\overline{\rho}_m)\,\tau^{\dagger\otimes 4}$.
Using Eq.~\eqref{eq:twirl_general}, we have
\begin{equation}
x_{m+1}
=
d\,\alpha(x_m)\Tr(Q\Pi_{\rm sym})
+ d\,\beta(x_m)\Tr\!\left(Q\tau^{\otimes 4}Q\tau^{\dagger\otimes 4}\Pi_{\rm sym}\right).
\end{equation}
The trace $\Tr(Q\tau^{\otimes 4}Q\tau^{\dagger\otimes 4}\Pi_{\rm sym})$ can be extracted from the single-shot calculation.
Substituting $a$, $b$ from Eq.~\eqref{eq:ab_values} and $\Tr(Q\Pi_{\rm sym})=(d+1)(d+2)/6$ into Eq.~\eqref{eq:Em_Mlin_after_twirl}, we find
\begin{equation}
\mE_{\mC}\!\left[M_{\rm lin}\!\left(T^{\otimes N}\mC\ket{\Phi}\right)\right]
=
1 - \frac{4}{d+4}
- \frac{6d}{(d+1)(d+2)(d+4)}\,
\Tr\!\left(Q\tau^{\otimes 4}Q\tau^{\dagger\otimes 4}\Pi_{\rm sym}\right),
\end{equation}
where $\mE_{\mC}\!\left[M_{\rm lin}\!\left(T^{\otimes N}\mC\ket{\Phi}\right)\right]$ is already known in closed form from Eq.~\eqref{eq:final_result_Mlin_TCliff}.
Therefore,
\begin{equation}
\Tr\!\left(Q\tau^{\otimes 4}Q\tau^{\dagger\otimes 4}\Pi_{\rm sym}\right)
=
\frac{3^N+3}{6}.
\end{equation}
Substituting this together with Eq.~\eqref{eq:alpha_beta_x} and
$\Tr(Q\Pi_{\rm sym})=(d+1)(d+2)/6$, we find
\begin{equation} \label{eq:recursion}
x_{m+1}
=
\frac{4(d-x_m)}{(d-1)(d+4)}
+
\frac{d(3^N+3)\big[(d+3)x_m-4\big]}
{(d-1)(d+1)(d+2)(d+4)}.
\end{equation}
Collecting the terms proportional to $x_m$, we define
\begin{equation}
\lambda_N
:=
\frac{d(d+3)(3^N-1)-8}{(d-1)(d+1)(d+2)(d+4)}.
\label{eq:Lambda}
\end{equation}
Eq.~\eqref{eq:recursion} can then be rewritten as
\begin{equation} \label{eq:xm1}
x_{m+1}
=
\lambda_N x_m + \frac{4(1-\lambda_N)}{d+3}.
\end{equation}
For $m=0$, the initial state $\ket{\Phi}$ is a stabilizer state with $M_{\rm lin}=0$, so $x_0=1=(4+d-1)/(d+3)$.
Putting this initial condition into the recursion [Eq.~\eqref{eq:xm1}], we obtain 
\begin{equation}
x_m
=
\frac{4+(d-1)\,\lambda_N^m}{d+3}
\label{eq:xm_closed}
\end{equation}
for every integer $m\ge 0$.
Combining Eq.~\eqref{eq:xm_closed} with Eq.~\eqref{eq:xm_def}, we obtain
\begin{equation}
\mE_{\mC_1,\dots,\mC_m}\!\left[M_{\rm lin}(U_m\ket{\Phi})\right]
= 1 - x_m
=
\frac{d-1}{d+3}\left(1-\lambda_N^m\right).
\label{eq:Mlin_multi}
\end{equation}
The long-time limit $x_m\to 4/(d+3)$ corresponds to
$M_{\rm lin}^{\rm Haar}=(d-1)/(d+3)$, i.e.,
$M_2^{\rm Haar}=\log_2(d+3)-2$.

In the large-$N$ limit, the leading term of $\lambda_N$ in Eq.~\eqref{eq:Lambda} is
\begin{equation}
\lambda_N = (3/4)^N + \mO\!\left((3/8)^N\right).
\label{eq:Lambda_asymptotic}
\end{equation}
If the fluctuation of $M_{\rm lin}$ is small, we can approximate
$\mE[M_2] \approx -\log_2(1-\mE[M_{\rm lin}]) = \log_2(d+3) - \log_2(4+(d-1)\lambda_N^m)$,
so that the gap to the Haar value is
\begin{equation}
\Delta M_2 := M_2^{\rm Haar} - \mE[M_2]
\approx
\log_2\!\left(1+\frac{d-1}{4}\lambda_N^m\right).
\end{equation}
In the late-time regime where $(d-1)\lambda_N^m/4 \ll 1$, this reduces to
\begin{equation}
\Delta M_2 \approx \frac{d-1}{4\ln 2}\lambda_N^m \propto \lambda_N^m.
\end{equation}
Since the total number of $T$ gates is $N_T = mN$ and $\lambda_N \sim (3/4)^N$ for large $N$, we have
\begin{equation}
\lambda_N^m \sim e^{-N\ln(4/3)\cdot m} = e^{-\ln(4/3)\,N_T}.
\end{equation}
Therefore the asymptotic decay rate per $T$ gate is
\begin{equation}
s_\infty = \ln(4/3),
\label{eq:s_infty}
\end{equation}
as reported in the main text.

\section{Convergence of $M_2$ to the Haar value at additional scrambling ratios and system sizes}
\label{sec:t-gates-extra-checks}

The main text demonstrates that the late-time gap $\Delta M_2 := M_2^{\mathrm{Haar}} - \mE[M_2]$ decays exponentially as $\ln\Delta M_2 \approx -s(\eta)\,N_T + b$, where $N_T$ is the total number of injected $T$ gates.
Here, we present two additional numerical simulations.

\begin{figure}[t]
\centering
\includegraphics[width=\textwidth]{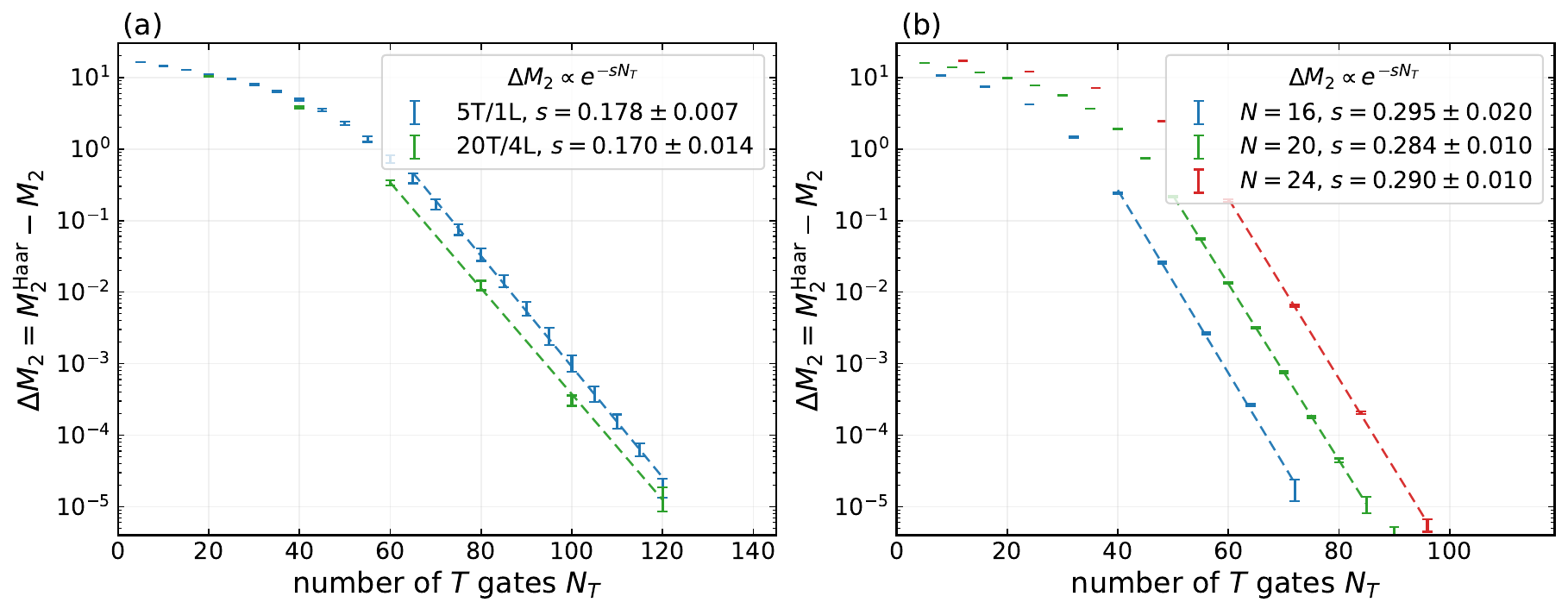}
\caption{
Convergence of $M_2$ to the Haar value: the decay rate $s$ in $\ln\Delta M_2 \approx -s\,N_T + b$.
(a)~Two $T$-gate schedules (``5T/1L'' and ``20T/4L'') at fixed $\eta=2$ and $N=20$.
(b)~Three system sizes ($N=16$, $20$, $24$) at fixed $\eta=5$.
Dashed lines denote linear fits in the late-time regime.
Each data point is averaged over $80$ random realizations; for $N\le 16$, $M_2$ is computed exactly, while for $N>16$ it is estimated with $2\times 10^4$ MC samples.
The decay rate $s$ and its error bar are obtained by bootstrap resampling over the realizations.
}
\label{fig:sm-decay-checks}
\end{figure}

Following the notation of the main text, ``$k_T$T/$k_L$L'' denotes a schedule in which $k_T$ parallel $T$ gates are applied after every $k_L$ Clifford layers.
Figure~\ref{fig:sm-decay-checks}(a) compares the ``5T/1L'' (uniform) and ``20T/4L'' (bursty) schedules at $N=20$.
Since one Clifford layer contains $N/2=10$ two-qubit Clifford gates, both schedules correspond to $\eta=2$.
The fitted slopes, $s=0.178\pm0.007$ and $s=0.170\pm0.014$ (uncertainty evaluated by bootstrap resampling), agree with each other, confirming that $s$ depends only on $\eta$ and not on the temporal distribution of $T$ gates.
Meanwhile, $\Delta M_2$ is smaller (corresponding to larger $M_2$) for the burstier ``20T/4L'' schedule, consistent with the observation in the main text that concentrating $T$ gates is slightly more efficient for generating nonstabilizerness.

Figure~\ref{fig:sm-decay-checks}(b) compares three system sizes $N=16$, $20$, and $24$ at fixed $\eta=5$.
Each cycle consists of $N_C=10$ Clifford layers followed by $N$ parallel $T$ gates (i.e., the ``$N$T/10L'' schedule), giving $\eta = 10/2 = 5$.
The fitted slopes, $s_{N=16}=0.295\pm0.020$, $s_{N=20}=0.284\pm0.010$, and $s_{N=24}=0.290\pm0.010$, are all very close to the analytical asymptote $s_\infty=\ln(4/3)\approx 0.288$, confirming that $s(\eta)$ is already close to saturation at $\eta=5$ for all three system sizes.

\section{Nonstabilizerness power of extensive Haar-random two-qubit gates}

We keep the setup of Sec.~\ref{sec: T gates} up to the Clifford-scrambled stabilizer state $\ket{\psi(N_C)}=\mC\ket{\Phi}$, but now replace the final $T^{\otimes N}$ layer by a layer of $N/2$ independent Haar-random two-qubit gates on neighboring pairs.
Assuming $N$ is even, we write
\begin{equation}
U:=\bigotimes_{j=1}^{N/2} U_j,
\end{equation}
where each $U_j$ acts on qubits $(2j-1,2j)$ and is drawn independently from the Haar measure on $U(4)$.
We study the doubly averaged quantity
\begin{equation}
\mathbb{E}_{U}\mathbb{E}_{\mC}\!\left[M_{\rm lin}\!\left(U\ket{\psi(N_C)}\right)\right],
\end{equation}
where $\mathbb{E}_{U}[\cdot]$ denotes the average over the final Haar-random two-qubit layer.
Repeating the steps leading to Eq.~\eqref{eq:Em_Mlin_after_twirl}, we obtain
\begin{equation}
\mathbb{E}_{U}\mathbb{E}_{\mC}\!\left[M_{\rm lin}\right]
=1-d\left[
a\,\Tr\!\left(Q\Pi_{\rm sym}\right)
+b\,\mathbb{E}_{U}\Tr\!\left(Q\,U^{\otimes 4} Q\,U^{\dagger \otimes 4}\,\Pi_{\rm sym}\right)
\right].
\label{eq:Em_Mlin_after_twirl_haar2q}
\end{equation}
Here $a$ and $b$ are given in Eq.~\eqref{eq:ab_values}, while $Q$ and $\Pi_{\rm sym}$ are defined in Sec.~\ref{sec: T gates}.
We decompose
\begin{equation}
Q=q_2^{\otimes N/2},
\qquad
q_2:=q\otimes q,
\qquad
q=\frac{1}{4}\Big(I^{\otimes4}+X^{\otimes4}+Y^{\otimes4}+Z^{\otimes4}\Big).
\end{equation}
Similarly, permutation operators decompose as
$W_\pi=\big(W_\pi^{(2)}\big)^{\otimes N/2}$,
where $W_\pi^{(2)}$ acts on the four-copy Hilbert space of a single qubit pair.
Since the final layer factorizes as $U=\bigotimes_{j=1}^{N/2}U_j$ with independent identically distributed $U_j$, we obtain
\begin{equation}
\mathbb{E}_{U}\Tr\!\left(Q\,U^{\otimes 4} Q\,U^{\dagger \otimes 4}\,\Pi_{\rm sym}\right)
=\frac{1}{24}\sum_{\pi\in S_4}
\left[
\tau_\pi
\right]^{N/2}.
\label{eq:trace_QUQ_haar2q}
\end{equation}
where
\begin{equation}
\tau_\pi
:=
\mathbb{E}_{u}\Tr\!\left(
u^{\otimes4}\,q_2\,u^{\dagger\otimes4}\,q_2\,W_\pi^{(2)}
\right),
\label{eq:tau_pi_haar2q}
\end{equation}
and $\mathbb{E}_{u}[\cdot]$ denotes the Haar average over a single two-qubit unitary $u\in U(4)$.
A straightforward two-qubit Haar-twirl calculation gives
\begin{equation}
\tau_e=\tau_{(12)(34)}=\frac{59}{14},
\qquad
\tau_{(12)}=\tau_{(1234)}=-\frac{2}{7},
\qquad
\tau_{(123)}=\frac{13}{28},
\end{equation}
so grouping permutations by cycle type yields
\begin{equation}
\mathbb{E}_{U}\Tr\!\left(Q\,U^{\otimes 4} Q\,U^{\dagger \otimes 4}\,\Pi_{\rm sym}\right)
=
\frac{1}{6}\left(\frac{59}{14}\right)^{N/2}
+\frac{1}{2}\left(-\frac{2}{7}\right)^{N/2}
+\frac{1}{3}\left(\frac{13}{28}\right)^{N/2}.
\end{equation}
Moreover,
\begin{equation}
\Tr\!\left(Q\Pi_{\rm sym}\right)=\frac{(d+1)(d+2)}{6}.
\end{equation}
Substituting these expressions into Eq.~\eqref{eq:Em_Mlin_after_twirl_haar2q}, we obtain
\begin{equation}
\mathbb{E}_{U}\mathbb{E}_{\mC}\! \left[M_{\rm lin}\!\left(U\mC\ket{\Phi}\right) \right]
=
1-\frac{4}{d+4}
-\frac{d}{(d+1)(d+2)(d+4)}
\left[
\left(\frac{59}{14}\right)^{N/2}
+3\left(-\frac{2}{7}\right)^{N/2}
+2\left(\frac{13}{28}\right)^{N/2}
\right],
\end{equation}
whose large-$N$ asymptotic form is
\begin{equation}
\mathbb{E}_{U}\mathbb{E}_{\mC}\! \left[M_{\rm lin}\!\left(U\mC\ket{\Phi}\right) \right]
=
1-\left(\frac{59}{224}\right)^{N/2}
-\frac{4}{2^N}
+\mathcal{O}\!\left(\left(\frac{13}{112}\right)^{N/2}\right).
\label{eq:final_result_Mlin_haar2qCliff}
\end{equation}
If the fluctuation of $M_{\rm lin}$ is small, we can approximate
\begin{equation}
\mathbb{E}_{U}\mathbb{E}_{\mC}[M_2]
\approx
-\log_2\!\left(
1-\mathbb{E}_{U}\mathbb{E}_{\mC}[M_{\rm lin}]
\right)
=
\frac{N}{2}\log_2\!\left(\frac{224}{59}\right)
+
\mathcal{O}\!\left(\left(\frac{56}{59}\right)^{N/2}\right)
=
0.962\ldots\,N
+\mathcal{O}\!\left((0.974\ldots)^N\right),
\end{equation}
which shows that one extensive layer of Haar-random two-qubit gates generates extensive nonstabilizerness on top of a Clifford-scrambled stabilizer state.
For the system sizes used in the main text, evaluating
$-\log_2\!\left(1-\mathbb{E}_{U}\mathbb{E}_{\mC}[M_{\rm lin}]\right)$
with the exact expression above gives
\begin{equation}
\frac{\mathbb{E}_{U}\mathbb{E}_{\mC}[M_2]}{N}
\approx
0.799 \ \ (N=12),\qquad
0.846 \ \ (N=16),\qquad
0.875 \ \ (N=20),
\end{equation}
in good agreement with the saturation values shown in the main text.

\section{Stabilizer R\'enyi entropy of mixed states} %
Let $\rho_A$ be the reduced density matrix of $|\psi\rangle\!\langle\psi|$ on a subsystem $A$ with $N_A$ qubits ($d_A=2^{N_A}$).
Its stabilizer R\'enyi entropy is defined as~\cite{leone2022}
\begin{equation}
M_\alpha(\rho_A)
:=\frac{1}{1-\alpha}\log_2\!\left(\sum_{P\in\mathcal{P}_A}\frac{|\Tr(\rho_A P)|^{2\alpha}}{d_A^{\alpha}}\right)-N_A-S_2(\rho_A),
\label{eq:Msre-mixed}
\end{equation}
where $\mathcal{P}_A$ denotes Pauli strings supported on $A$.
Without loss of generality, we take $A$ to be the first $N_A$ qubits and write
$x=(x_A,0)$ and $z=(z_A,0)$, where
$x_A,z_A\in\mF_2^{N_A}$ and the remaining $N-N_A$ entries are zero.
Then,
\begin{equation}
\mathcal{P}_A
:=
\left\{
P_{x,z}\;\middle|\;
x=(x_A,0),\,
z=(z_A,0),\,
x_A,z_A\in\mF_2^{N_A} \, .
\right\}
\end{equation}
For $P\in\mathcal{P}_A$, we have
\begin{equation}
\Tr(\rho_A P)=\langle\psi|P|\psi\rangle=:c_P.
\end{equation}
The second R\'enyi entropy satisfies
\begin{equation}
S_2(\rho_A)=-\log_2\!\left(\sum_{P\in\mathcal{P}_A}|c_P|^2/d_A\right).
\end{equation}
Therefore, evaluating $M_\alpha(\rho_A)$ only requires Pauli correlators supported on $A$.
Write a computational-basis configuration as $b=(a,e)$, where
$a\in\mF_2^{N_A}$ labels subsystem $A$ and
$e\in\mF_2^{N-N_A}$ labels its complement.
For a Pauli string supported on $A$, namely
$P=P_{(x_A,0),(z_A,0)}$ with
$x_A,z_A\in\mF_2^{N_A}$, we have
\begin{equation}
\langle\psi|P|\psi\rangle
=
e^{i\phi}
\sum_{a,e}
\overline{\psi(a,e)}\,
(-1)^{z_A\cdot a}\,
\psi(a\oplus x_A,e),
\end{equation}
where the phase $e^{i\phi}$ is irrelevant for $|\langle\psi|P|\psi\rangle|$.
Defining
\begin{equation}
h_{x_A}(a)
:=
\sum_{e\in\mF_2^{N-N_A}}
\overline{\psi(a,e)}\,\psi(a\oplus x_A,e),
\end{equation}
we obtain
\begin{equation}
\langle\psi|P_{(x_A,0),(z_A,0)}|\psi\rangle
=
e^{i\phi}
\sum_{a\in\mF_2^{N_A}}
(-1)^{z_A\cdot a}\,
h_{x_A}(a).
\end{equation}
Thus, for each fixed $x_A$, the full set of correlators
$\{\langle\psi|P_{(x_A,0),(z_A,0)}|\psi\rangle\}_{z_A\in\mF_2^{N_A}}$
is obtained by an $N_A$-qubit FWHT of the length-$2^{N_A}$ vector $h_{x_A}(a)$.
Computing $h_{x_A}(a)$ involves the sum over $e\in\mF_2^{N-N_A}$, costing $\mO(2^{N-N_A})$, so computing $h_{x_A}(a)$ for all $a$ costs $\mO(2^N)$.
The subsequent FWHT costs $\mO(N_A2^{N_A})$.
Sweeping over all $x_A\in\mF_2^{N_A}$ therefore yields the total cost
\begin{equation}
\mO\!\left(2^{N_A}\bigl(2^N+N_A2^{N_A}\bigr)\right)
=
\mO\!\left(2^{N+N_A}+N_A2^{2N_A}\right).
\end{equation}
As a comparison, if we use brute-force enumeration to compute $\langle\psi|P|\psi\rangle$ for each $P\in\mathcal{P}_A$, the cost is $\mO(2^{N+2N_A})$.

\bibliography{ref_v2}